\documentclass[aps,prc,preprint,amsmath,superscriptaddress,floatfix,nofootinbib]{revtex4-1}
\usepackage{graphicx,epsfig}
\usepackage{color}
\newcommand{\beqy}{\begin{eqnarray}}
\newcommand{\eeqy}{\end{eqnarray}}

\begin{document}

\title{Entrainment effects in neutron-proton mixtures within the nuclear-energy density functional theory. I. Low-temperature limit.}

 \author{N. Chamel}
 \affiliation{Institute of Astronomy and Astrophysics, Universit\'e Libre de Bruxelles, CP 226, Boulevard du Triomphe, B-1050 Brussels, Belgium}
  \author{V. Allard}
 \affiliation{Institute of Astronomy and Astrophysics, Universit\'e Libre de Bruxelles, CP 226, Boulevard du Triomphe, B-1050 Brussels, Belgium}

\date{\today}

\begin{abstract}
Mutual entrainment effects in cold neutron-proton mixtures are studied in the framework of the 
self-consistent nuclear energy-density functional theory. Exact expressions for the mass 
currents, valid for both homogeneous and inhomogeneous systems, are directly derived from the 
time-dependent Hartree-Fock equations with no further approximation. The equivalence with the 
Fermi-liquid expression is also demonstrated. Focusing on neutron-star 
cores, a convenient and simple analytical formulation of the entrainment matrix in terms of the 
isovector effective mass is found, thus allowing to relate entrainment phenomena in neutron stars 
to isovector giant dipole resonances in finite nuclei. Results obtained with different functionals 
are presented. These include the Brussels-Montreal functionals, for which unified equations of state 
of neutron stars have been recently calculated. 
\end{abstract}

\maketitle

\section{Introduction}

Neutron stars are unique celestial bodies in that their core is expected to contain neutron and proton superfluids, the former permeating also the inner part of the crust~\cite{ginzburg1965,wolf1966,tamagaki1970,hoffberg1970,takatsuka1972,amundsen1985,ainsworth1989} 
(see, e.g., Refs.~\cite{lombardo2001,dean2003,baldo2012,gandolfi2015} for reviews). Predicted before the actual discovery of these compact stars~\cite{migdal1959}, nuclear superfluidity has found strong  support from observations of pulsar frequency glitches~\cite{alpar1985,haskell2015}, and more recently from the rapid cooling of the young neutron star in Cassiopeia A supernova remnant~\cite{page2011,shternin2011,wijngaarden2019} (but see also Ref.~\cite{posselt2018}). Superfluidity in neutron stars may leave its imprints on other astrophysical phenomena (see, e.g., Refs.~\cite{chamel2017,haskell2018}).

Although superfluid neutrons and protons in a cold mature neutron star can flow with different velocities, their dynamics are not completely independent from each other. Despite the absence of viscous drag, the neutron superfluid in the crust does not flow freely due to scattering by inhomogeneities. The neutron superfluid is thus effectively entrained by the crust (see, e.g. Ref.~\cite{chamel2017b} for a recent review). 
Likewise, neutrons and protons in the core are mutually coupled by nondissipative entrainment effects of the kind originally discussed by Andreev and Bashkin in the context of superfluid $^{4}$He-$^{3}$He mixtures~\cite{andreev1976}: the mass current $\pmb{\rho_q}$ of one nucleon species ($q=n,p$ for neutron, proton respectively) is found to depend on the superfluid velocities $\pmb{V_q}$ of both species, i.e. 
\beqy\label{eq:def-matrix}
\pmb{\rho_q} =\sum_{q^\prime} \rho_{qq^\prime} \pmb{V_{q^\prime}}\, .
\eeqy
These effects may have important consequences for the global dynamics of a neutron star. For instance, electron scattering off the magnetic field induced by the circulation of entrained protons around individual neutron superfluid vortices leads to a very strong frictional coupling between the neutron superfluid in the core and the electrically charged particles~\cite{alpar1984}. The (symmetric) entrainment matrix $\rho_{q
q^\prime}$ in neutron-proton mixtures has been previously calculated in the framework of the Fermi liquid theory~\cite{borumand1996,gusakov2005,chamel2006,gusakov2009,gusakov2009b,leinson2017,leinson2018}. An alternative approach based on relativistic mean-field models has been followed in Refs.~\cite{comer2003,kheto2014,sourie2016}. 

In this paper, entrainment effects are studied within the self-consistent nuclear energy-density functional theory. In Section~\ref{sec:micro-currents}, we derive  the  microscopic expressions for the neutron and proton mass currents in the framework of the time-dependent Hartree-Fock (TDHF) method (see, e.g. Refs.~\cite{simenel2018,stevenson2019} for recent reviews). Applications to neutron-star cores are discussed in Section~\ref{sec:core}, where the entrainment matrix is calculated. The equivalence with the Fermi-liquid expression obtained earlier is explicitly demonstrated. Numerical results are presented for extended Skyrme functionals, for which unified equations of state of neutron stars have been recently calculated~\cite{potekhin2013,pearson2018}. Other functionals are also considered for comparison.  

\section{Microscopic expressions of the mass currents}
\label{sec:micro-currents}

In the following, we will consider cold neutron-proton mixtures at temperatures $T$ much lower than the critical  temperatures of nuclear superfluidity. We shall further suppose that currents are small compared to the critical currents for the breakdown of nuclear superfluidity. With these assumptions, the influence of nuclear pairing on the entrainment matrix can be safely ignored (see e.g. Ref.~\cite{leinson2017}).

\subsection{Time-dependent Hartree-Fock equations}

The total energy $E$ of a nucleon-matter element of volume $V$ is supposed to be a functional of the following local densities and currents:

\noindent (i) the nucleon number density at position $\pmb{r}$ and time $t$
\beqy
\label{eq:dens-def}
n_q(\pmb{r},t) = \sum_{\sigma=\pm 1}n_q(\pmb{r}, \sigma; \pmb{r}, \sigma;t)\, ,
\eeqy
(ii) the kinetic density at position $\pmb{r}$ and time $t$
\beqy
\label{eq:kin-def}
\tau_q(\pmb{r},t) = \sum_{\sigma=\pm 1}\int\,{\rm d}^3\pmb{r^\prime}\,\delta(\pmb{r}-\pmb{r^\prime}) \pmb{\nabla}\cdot\pmb{\nabla^\prime}
n_q(\pmb{r}, \sigma; \pmb{r^\prime}, \sigma;t)\, ,
\eeqy
(iii) and the momentum density (in units of $\hbar$) at position $\pmb{r}$ and time $t$
\beqy
\label{eq:mom-def}
\pmb{j_q}(\pmb{r},t)=-\frac{\rm i}{2}\sum_{\sigma=\pm 1}\int\,{\rm d}^3\pmb{r^\prime}\,\delta(\pmb{r}-\pmb{r^\prime}) (\pmb{\nabla} -\pmb{\nabla^\prime})n_q(\pmb{r}, \sigma; \pmb{r^\prime}, \sigma;t)\, ,
\eeqy
where $n_q(\pmb{r}, \sigma; \pmb{r^\prime}, \sigma^\prime;t)$ is the density matrix in coordinate space 
(denoting the spin states by $\sigma,\sigma^\prime$; allowed values are $1,-1$ for spin up, spin down respectively). We consider here pure nucleon states as in most existing functionals. The more general formalism involving isospin mixing has been developed in Ref.~\cite{perlinska2004}. 

The dynamics of the system is governed by the TDHF equations, which are generally written in a basis of discrete single-particle states (labelled by $i$, $j$, etc.)  as~\cite{ringschuck1980} 
\beqy
\label{eq:TDHF-trad}
\mathrm{i}\hbar \frac{\partial n_q^{ij}}{\partial t}=\sum_k (h_q^{ik}n_q^{kj}-n_q^{ik}h_q^{kj}) \, ,
\eeqy
where the (Hermitian) Hamiltonian matrix $h_q^{ij}$ is defined by 
\beqy
h_q^{ij}=\frac{\partial E}{\partial n_q^{ji}}=(h_q^{ji})^*
\eeqy
(the symbol $*$ denoting complex conjugation). 

As shown in Appendix~\ref{app:TDHF}, the TDHF equations can be equivalently expressed in coordinate space as
\beqy
\label{eq:TDHF}
\mathrm{i}\hbar \frac{\partial n_q(\pmb{r}, \sigma; \pmb{r^\prime}, \sigma^\prime;t)}{\partial t}=h_q(\pmb{r},t)n_q(\pmb{r}, \sigma; \pmb{r^\prime}, \sigma^\prime;t) -  h_q(\pmb{r^{\prime}},t)^*\, n_q(\pmb{r}, \sigma; \pmb{r^{\prime}}, \sigma^\prime;t) \, .
\eeqy
in which the single-particle Hamiltonian $h_q$ is given by 
\beqy
\label{eq:Hamiltonian}
h_q(\pmb{r},t)&=&-\pmb{\nabla}\cdot \frac{\hbar^2}{2 m_q^\oplus(\pmb{r},t)}\pmb{\nabla} + U_q(\pmb{r},t)-
\frac{\rm i}{2}\biggl[\pmb{I_q}(\pmb{r},t)\cdot\pmb{\nabla}+\pmb{\nabla}\cdot\pmb{I_q}(\pmb{r},t)\biggr]
\eeqy
with the various fields defined by the functional derivatives of the energy 
\beqy
\label{eq:def-fields}
\frac{\hbar^2}{2 m_q^\oplus(\pmb{r},t)}=\frac{\delta E}{\delta \tau_q(\pmb{r},t)}, \ U_q(\pmb{r},t)=\frac{\delta E}{\delta n_q(\pmb{r},t)},\ 
\pmb{I_q}(\pmb{r},t)=  \frac{\delta E}{\delta \pmb{j_q}(\pmb{r},t)}\, .
\eeqy

\subsection{Mass currents, velocities and momenta}

Due to neutron-proton interactions, the nucleon mass current $\pmb{\rho_q}$ is not simply given by the momentum density $\hbar\pmb{j_q}$. The mass current can be rigorously calculated from the TDHF matrix equations~(\ref{eq:TDHF}),  whose diagonal part 
can be rearranged in the form of continuity equations for nucleons of type $q$ after summing over spins following the seminal work of Ref.~\cite{engel1975}
\beqy
 \frac{\partial \rho_q(\pmb{r},t)}{\partial t}+\pmb{\nabla}\cdot\pmb{\rho_q}(\pmb{r},t) = 0\, . 
\eeqy
Using the Hamiltonian (\ref{eq:Hamiltonian}) and the definitions~(\ref{eq:dens-def})-(\ref{eq:mom-def}), 
we thus find 
\beqy\label{eq:mass-current}
\pmb{\rho_q}(\pmb{r},t) =\frac{m}{m_q^\oplus(\pmb{r},t)} \hbar \pmb{j_q}(\pmb{r},t)+\rho_q(\pmb{r},t) \frac{\pmb{I_q}(\pmb{r},t)}{\hbar}\, ,
\eeqy
where $m$ denotes the nucleon mass, ignoring the small difference between the neutron and proton masses. 

The energy $E$ of a nucleon matter element can be decomposed into a kinetic term
\beqy
E_{\rm kin}=\int {\rm d}^3\pmb{r}\, \frac{\hbar^2}{2m}\tau(\pmb{r},t)\, , 
\eeqy
where $\tau=\tau_n+\tau_p$, 
a Coulomb term $E_{\rm Coul}$ and a nuclear term $E_{\rm nuc}$, i.e. 
\beqy
\label{eq:energy}
E=E_{\rm kin}+E_{\rm Coul}+E_{\rm nuc}\quad .
\eeqy
Assuming nuclear isospin symmetry, $E_{\rm nuc}$ remains unaffected if neutron and proton densities and currents are interchanged. It is convenient to introduce an isospin index equal to $0$ for isoscalar quantities and $1$ for isovector quantities. The former (also written without any subscript) are sums over neutrons and protons (e.g. $n_0\equiv n=n_n+n_p$) while the latter are differences between neutrons and protons (e.g. $n_1=n_n-n_p$). Due to Galilean invariance, the nuclear-energy terms contributing to the mass currents, denoted by $E^j_{\rm nuc}$, can only depend on the combinations $X_0(\pmb{r},t)=n_0(\pmb{r},t)\tau_0(\pmb{r},t) -  j_0(\pmb{r},t)^2$ and $X_1(\pmb{r},t)=n_1(\pmb{r},t)\tau_1(\pmb{r},t) -  j_1(\pmb{r},t)^2$, as shown, e.g.,   Ref.~\cite{dobaczewski1995}. Therefore, the functional derivatives of $E^j_{\rm nuc}$ with respect to $\tau_q(\pmb{r},t)$ and $j_q(\pmb{r},t)$ can be written as
\beqy\label{eq:def-Bq}
\frac{\delta E^j_{\rm nuc}}{\delta \tau_q(\pmb{r},t)}=\frac{\hbar^2}{2 m_q^\oplus(\pmb{r},t)}-\frac{\hbar^2}{2m} = \Biggl[\frac{\delta E^j_{\rm nuc}}{\delta X_0(\pmb{r},t)}-\frac{\delta E^j_{\rm nuc}}{\delta X_1(\pmb{r},t)}\Biggr]n + 2 n_q \frac{\delta E^j_{\rm nuc}}{\delta X_1(\pmb{r},t)}
\, ,
\eeqy
\beqy\label{eq:def-Iq}
\frac{\delta E^j_{\rm nuc}}{\delta \pmb{j_q}(\pmb{r},t)}=\pmb{I_q}(\pmb{r},t)=-2 \pmb{j} \Biggl[\frac{\delta E^j_{\rm nuc}}{\delta X_0(\pmb{r},t)}-\frac{\delta E^j_{\rm nuc}}{\delta X_1(\pmb{r},t)}\Biggr]  -4 \pmb{j_q} \frac{\delta E^j_{\rm nuc}}{\delta X_1(\pmb{r},t)}
\, .
\eeqy
Using Eqs.~(\ref{eq:def-Bq}) and (\ref{eq:def-Iq}), the mass current~(\ref{eq:mass-current}) can be expressed in terms of the momentum densities only as 
\beqy\label{eq:mass-current-C}
\pmb{\rho_q}(\pmb{r},t) &=&\hbar \pmb{j_q}(\pmb{r},t)\Biggl\{1+\frac{2}{\hbar^2}\Biggl[\frac{\delta E^j_{\rm nuc}}{\delta X_0(\pmb{r},t)}-\frac{\delta E^j_{\rm nuc}}{\delta X_1(\pmb{r},t)}\Biggr]\rho(\pmb{r},t)\Biggr\}\nonumber \\ 
&&-\hbar \pmb{j}(\pmb{r},t) \frac{2}{\hbar^2}\Biggl[\frac{\delta E^j_{\rm nuc}}{\delta X_0(\pmb{r},t)}-\frac{\delta E^j_{\rm nuc}}{\delta X_1(\pmb{r},t)}\Biggr]\rho_q(\pmb{r},t)\, .
\eeqy 
While neutron and proton mass currents are not separately aligned with their corresponding momenta, it can be easily seen that the total mass current coincides with the total momentum density 
\beqy\label{eq:Galilean}
\pmb{\rho}(\pmb{r},t)=\pmb{\rho_n}(\pmb{r},t)+\pmb{\rho_p}(\pmb{r},t)=\hbar \pmb{j}(\pmb{r},t)\, .
\eeqy

The mean mass current $\pmb{\rho_q}(t)$ in the volume $V$ is obtained by integrating the corresponding local current $\pmb{\rho_q}(\pmb{r},t)$. Decomposing the density matrix in a single-particle basis~(\ref{eq:dens-matrix}) using Eqs.~(\ref{eq:dens-def}), (\ref{eq:mom-def}), and (\ref{eq:mass-current}),  the mean mass current can thus be written as
\beqy\label{eq:mean-current}
\pmb{\rho_q}(t) =\frac{1}{V}\int\mathrm{d}^3\pmb{r}\,\pmb{\rho_q}(\pmb{r},t)= \frac{ m  }{V} \sum_{i,j}\  n_q^{ij}\, \pmb{v_{ji}^{(q)}}\, ,
\eeqy
where 
\beqy
\pmb{v_{ji}^{(q)}}=\sum_\sigma\int\mathrm{d}^3\pmb{r}\,  \varphi_{j}^{(q)}(\pmb{r},\sigma)^* \pmb{v^{(q)}}(\pmb{r},t)\varphi_{i}^{(q)}(\pmb{r},\sigma) 
\eeqy
are the matrix elements of the velocity operator
\beqy
\pmb{v^{(q)}}(\pmb{r},t) =\frac{-{\rm i} \hbar}{2}\left[\frac{1}{m_q^\oplus(\pmb{r},t)}\pmb{\nabla}+\pmb{\nabla}\frac{1}{m_q^\oplus(\pmb{r},t)}\right]+\frac{1}{\hbar}\pmb{I_q}(\pmb{r},t)\, .
\eeqy
That $\pmb{v^{(q)}}(\pmb{r},t)$ is a velocity operator is confirmed by the application of the Ehrenfest theorem (see, e.g. Ref.~\cite{messiah2014})
\beqy\label{eq:velocity-op}
\pmb{v_{ji}^{(q)}}=\sum_\sigma\int\mathrm{d}^3\pmb{r}\,  \varphi_{j}^{(q)}(\pmb{r},\sigma)^*
\frac{1}{\mathrm{i} \hbar }\biggl[\pmb{r} h_q(\pmb{r},t) - h_q(\pmb{r},t)\pmb{r} \biggr]\varphi_{i}^{(q)}(\pmb{r},\sigma) 
\, .
\eeqy
In the canonical basis for which the density matrix is diagonal, i.e. $n_q^{ij}=\widetilde{n}^{(q)}_i \delta_{ij}$ where $\widetilde{n}^{(q)}_i$ represents the occupation number of the single-particle state $i$ ($\delta_{ij}$ being the Kronecker symbol), the mean mass current takes a particularly simple form 
\beqy\label{eq:mean-current-canonical}
\pmb{\rho_q} = \frac{ m  }{V} \sum_i\  \widetilde{n}^{(q)}_i\, \pmb{v_i^{(q)}}\, ,
\eeqy
with $\pmb{v_i^{(q)}}\equiv \pmb{v_{ii}^{(q)}}$ denoting the mean velocity of the state $i$. 

The equations derived so far for the mass currents are very general since we only made use of the TDHF equations~(\ref{eq:TDHF}) with no further approximation. In particular, Eqs.~(\ref{eq:mass-current}),  (\ref{eq:mass-current-C}) and (\ref{eq:mean-current-canonical}) are applicable to both homogeneous and inhomogeneous systems such as the core and the crust of a neutron star respectively. 

\subsection{Relation to the Fermi liquid theory}

In systems that have some translational symmetry (but not necessarily homogeneous), any single-particle state can be labelled by a wave vector $\pmb{k}$. Assuming further that the system is stationary, the TDHF equation~(\ref{eq:TDHF-trad}) shows that the Hamiltonian and density matrices commute, and therefore share the same eigenstates. In other words, the single-particle Hamiltonian is diagonal in the canonical basis
\beqy
\label{eq:HF}
h_q(\pmb{r})\varphi_{\pmb{k}}^{(q)}(\pmb{r},\sigma)=\varepsilon_{\pmb{k}}^{(q)}\varphi_{\pmb{k}}^{(q)}(\pmb{r},\sigma)\, .
\eeqy
As shown in Appendix~\ref{app:group-vel}, the mean velocity $\pmb{v_k^{(q)}}$ of a state $\pmb{k}$  can be expressed as 
\beqy\label{eq:group-vel}
\pmb{v_k^{(q)}}=\frac{1}{\hbar}\pmb{\nabla_k} \varepsilon_{\pmb{k}}^{(q)}\, .
\eeqy
The mean current is thus given by the familiar expression
\beqy\label{eq:mass-current-def}
\pmb{\rho_q} = \frac{ m  }{V} \sum_{\pmb{k}}\  \widetilde{n}^{(q)}_{\pmb{k}}\, \pmb{v_k^{(q)}}\, .
\eeqy
This demonstrates the equivalence between the definition of the mass currents in the Fermi liquid theory, namely Eqs.~(\ref{eq:group-vel}) and (\ref{eq:mass-current-def}), and the expression~(\ref{eq:mean-current}) derived from the TDHF equations~(\ref{eq:TDHF}).

\section{Entrainment effects in neutron-star cores}
\label{sec:core}

We focus here on homogeneous nucleon matter with stationary currents. All fields are therefore spatially uniform and independent of time. 

\subsection{Andreev-Bashkin matrix in the Fermi liquid theory}

The entrainment matrix was previously calculated in the framework of the Fermi liquid theory by considering small perturbations of the static ground-state configuration~\cite{borumand1996}. In the  presence of currents, the neutron and proton Fermi surfaces are shifted by a vector $\pmb{Q_n}$ and $\pmb{Q_p}$ respectively, which are related to the ``superfluid velocities'' by 
\beqy
\pmb{V_q} =\frac{\hbar\pmb{Q_q}}{m}\, .
\eeqy
To first order in $Q_q/k_{\rm F}^{(q)}$, where $k_{\rm F}^{(q)}=(3\pi^2 n_q)^{1/3}$ denotes the Fermi wave number, the induced mass current, 
\beqy\label{eq:mass-current-lin}
\pmb{\rho_q} \approx \delta \pmb{\rho_q} = \frac{ m  }{V} \sum_{\pmb{k}}\ ( \delta\widetilde{n}^{(q)}_{\pmb{k}}\, \pmb{v_k^{(q)}} + \widetilde{n}^{(q)}_{\pmb{k}}\, \delta\pmb{v_k^{(q)}})\, ,
\eeqy
can be written in the form of Eq.~(\ref{eq:def-matrix}) with the entrainment matrix~\cite{borumand1996}
\beqy\label{eq:entrainment-matrix-FL}
\rho_{qq^\prime}=\sqrt{\rho_q \rho_{q^\prime}}\frac{m}{\sqrt{m_q^\oplus m_{q^\prime}^\oplus}} \left(\delta_{qq^\prime}+\frac{\mathcal{F}_1^{q q^\prime}}{3}\right) \, ,
\eeqy
where $m_q^\oplus$ is the (Landau) effective mass and $\mathcal{F}_1^{q q^\prime}$ are dimensionless $\ell=1$ Landau parameters. 

\subsection{Andreev-Bashkin matrix in the TDHF theory}

As we will now show the entrainment matrix can be calculated \emph{exactly} in the TDHF theory. Introducing the ``superfluid velocity'' 
\beqy
\pmb{V_q} =\frac{\hbar}{\rho_q}  \pmb{j_q}\, ,
\eeqy
and using Eq.~(\ref{eq:mass-current-C}), the entrainment matrix is found to be given by 
\begin{align}
\label{eq:entrainment-matrix-TDHF}
\rho_{nn}&=\rho_n  \Biggl[1 + \frac{2}{\hbar^2}\left(\frac{\delta E^j_{\rm nuc}}{\delta X_0}-\frac{\delta E^j_{\rm nuc}}{\delta X_1}\right)\rho_{p}\Biggr] \nonumber \\ 
\rho_{pp}&=\rho_p  \Biggl[1 + \frac{2}{\hbar^2}\left(\frac{\delta E^j_{\rm nuc}}{\delta X_0}-\frac{\delta E^j_{\rm nuc}}{\delta X_1}\right)\rho_{n}\Biggr] \nonumber \\ 
\rho_{np}&=\rho_{pn}=-\rho_n \rho_p  \frac{2}{\hbar^2}\left(\frac{\delta E^j_{\rm nuc}}{\delta X_0}-\frac{\delta E^j_{\rm nuc}}{\delta X_1}\right) \, .
\end{align}
Let us stress that the functional derivatives of $E^j_{\rm nuc}$ may generally depend on the nucleon densities and currents unless $E^j_{\rm nuc}$ is a linear combination of $X_0$ and $X_1$ or the functional derivatives of $E^j_{\rm nuc}$ with respect to $X_0$ and $X_1$ cancel exactly. Unlike the Fermi-liquid expression~(\ref{eq:entrainment-matrix-FL}), the mass currents obtained from the TDHF expression~(\ref{eq:entrainment-matrix-TDHF}) may thus depend nonlinearly on the superfluid velocities. 

The Fermi-liquid expression~(\ref{eq:entrainment-matrix-FL}) is recovered by evaluating the functional derivatives of $E^j_{\rm nuc}$ with respect to $X_0$ and $X_1$ in the static configuration, i.e. by setting $\pmb{j_q}=\pmb{0}$ after derivation. To verify that Eq.~(\ref{eq:entrainment-matrix-TDHF}) reduces to (\ref{eq:entrainment-matrix-FL}), we need to calculate the Landau effective mass and the $\ell=1$ Landau parameters in the TDHF theory. It follows immediately from Eq.~(\ref{eq:s-p-energy}) that the Landau effective mass defined as (the subscript '0' indicating that the derivative is evaluated in the absence of currents with $\pmb{k}$ lying on the corresponding Fermi surface)
\beqy
\frac{1}{m_q^\oplus}=\frac{1}{\hbar^2 k_{\rm F}^{(q)}}\frac{d\varepsilon_{\pmb{k}}^{(q)}}{dk}\Biggr\vert_0
\eeqy
 coincides with the effective mass appearing in the TDHF theory. We have thus used the same symbol. The Landau parameters are obtained from the spin-averaged quasiparticle interaction defined by
\beqy\label{eq:q-p-interaction}
f^{q q^\prime}({\pmb k}, {\pmb k^\prime})=\frac{\delta^2 E}{\delta  \widetilde{n}^{(q)}_{\pmb{k}} \delta\widetilde{n}^{(q^\prime)}_{\pmb{k^\prime}}}\Biggr\vert_0=\frac{\delta \varepsilon_{\pmb{k}}^{(q)}}{\delta\widetilde{n}^{(q^\prime)}_{\pmb{k^\prime}}}\Biggr\vert_0\, .
\eeqy
The quasiparticle interaction is further expanded into Legendre polynomials 
\beqy
f^{q q^\prime}({\pmb  k}, {\pmb k^\prime}) = \sum_{\ell} f^{qq^\prime}_\ell P_\ell (\cos \theta)
\eeqy
where $\theta$ is the angle between the wave vectors $\pmb k$ and $\pmb k^\prime$ lying on the corresponding Fermi surface.  The dimensionless Landau parameters $\mathcal{F}_1^{q q^\prime}$ appearing in Eq.~(\ref{eq:entrainment-matrix-FL}) are defined by
\beqy
\mathcal{F}_1^{q q^\prime}=
\sqrt{{\cal N}_q {\cal N}_{q^\prime}} f^{qq^\prime}_\ell \, ,
\eeqy
in which ${\cal N}_q$ is the density of quasiparticle states at the Fermi surface,
\beqy
{\cal N}_q=\frac{m_q^\oplus k_{\rm F}^{(q)}}{\hbar^2 \pi^2}
\,.
\eeqy
In the TDHF theory for homogeneous matter (see Appendix~\ref{app:group-vel}), the quasiparticle energies are given by Eq.~(\ref{eq:s-p-energy}). From the general definition~(\ref{eq:q-p-interaction}), it follows that only the term $\pmb{k}\cdot\pmb{I_q}$ contributes to the $\ell=1$ Landau parameters. Using Eq.~(\ref{eq:def-Iq}) and remarking from Eq.~(\ref{eq:mom-def}) that the momentum density (in the canonical basis) reduces to 
\beqy
\pmb{j_q}=\sum_{\pmb{k}} \pmb{k}\, \widetilde{n}^{(q)}_{\pmb{k}}\, , 
\eeqy
the term $\pmb{k}\cdot\pmb{I_q}$ can be explicitly written as
\beqy
\pmb{k}\cdot\pmb{I_q} = -2 \sum_{\pmb{k^\prime}} \pmb{k}\cdot\pmb{k^\prime}\,  \widetilde{n}^{(q)}_{\pmb{k^\prime}} \Biggl[ \frac{\delta E^j_{\rm nuc}}{\delta X_0}+\frac{\delta E^j_{\rm nuc}}{\delta X_1}\Biggr]-2 \sum_{\pmb{k^\prime}} \pmb{k}\cdot\pmb{k^\prime}\,  \widetilde{n}^{(q^\prime)}_{\pmb{k^\prime}} \Biggl[ \frac{\delta E^j_{\rm nuc}}{\delta X_0}-\frac{\delta E^j_{\rm nuc}}{\delta X_1}\Biggr]\, .
\eeqy
The $\ell=1$ Landau parameters can be readily obtained by taking the derivatives of the above expression with respect to $\widetilde{n}^{(q)}_{\pmb{k^\prime}}$ and $\widetilde{n}^{(q^\prime)}_{\pmb{k^\prime}}$: 
\beqy
f_1^{qq}=-2\Biggl[ \frac{\delta E^j_{\rm nuc}}{\delta X_0}\biggr\vert_0+\frac{\delta E^j_{\rm nuc}}{\delta X_1}\biggr\vert_0\Biggr] (k_{\rm F}^{(q)})^2\, ,  
\eeqy
\beqy
f_1^{qq^\prime}=-2\Biggl[ \frac{\delta E^j_{\rm nuc}}{\delta X_0}\biggr\vert_0-\frac{\delta E^j_{\rm nuc}}{\delta X_1}\biggr\vert_0\Biggr] k_{\rm F}^{(q)}k_{\rm F}^{(q^\prime)}\, .
\eeqy 
Inserting the corresponding dimensionless parameters in Eq.~(\ref{eq:entrainment-matrix-FL}) leads to an  expression similar to Eq.~(\ref{eq:entrainment-matrix-TDHF}) except that the derivatives are now evaluated for vanishing currents. 

\subsection{Entrainment and isovector effective mass}

Due to Galilean invariance, as embedded in Eq.~(\ref{eq:Galilean}), it can be easily seen from Eq.~(\ref{eq:entrainment-matrix-TDHF}) that the entrainment matrix elements are not all independent but are related by the following identities
\beqy
\rho_{nn}+\rho_{np}=\rho_n \, , \ \rho_{pp}+\rho_{pn}=\rho_p \, .
\eeqy
This means that entrainment effects can be completely characterized by only one independent parameter, 
such as the dimensionless determinant of the entrainment matrix
\beqy
\Upsilon = \frac{\rho_{nn}\rho_{pp}-\rho_{np}^2}{\rho_n \rho_p}\, .
\eeqy
This parameter directly appears in the perturbed  hydrodynamical equations and 
is therefore important for the study of oscillation modes (see, e.g., Refs.~\cite{lindblom1994,andersson2001,lee2003}). Introducing the asymmetry parameter $\eta=(n_n-n_p)/n$, the entrainment matrix elements 
can thus be equivalently expressed as 
\beqy\label{eq:matrix-param1}
\rho_{nn}=\frac{1}{2}\rho\left(1+\eta\right)-\frac{1}{4}\rho\left(1-\eta^2\right)(1-\Upsilon)\, , 
\eeqy
\beqy\label{eq:matrix-param2}
\rho_{pp}=\frac{1}{2}\rho\left(1-\eta\right)-\frac{1}{4}\rho\left(1-\eta^2\right)(1-\Upsilon)\, , 
\eeqy
\beqy\label{eq:matrix-param3}
 \rho_{np}=\frac{1}{4}\rho\left(1-\eta^2\right)(1-\Upsilon)=\rho_{pn}\, .
\eeqy
The deviation of $\Upsilon$ from unity is a measure of the importance of entrainment effects. This parameter appears to have a simple physical meaning: it coincides with 
the inverse of the isovector effective mass defined by 
\beqy\label{eq:isoeffmass}
\frac{m}{m_v^\oplus}=\left(\frac{m}{m_n^\oplus}-\frac{n_n}{n_p}\frac{m}{m_p^\oplus}\right)\left(1-\frac{n_n}{n_p}\right)^{-1}\, .
\eeqy
Introducing the isoscalar effective mass
\beqy
\frac{m}{m_s^\oplus}=\frac{1}{2}\left(\frac{m}{m_n^\oplus}+\frac{m}{m_p^\oplus}\right)\, ,
\eeqy
the nucleon effective masses can be equivalently written as 
\beqy
\frac{m}{m_q^\oplus}=\frac{2n_q}{n}\frac{m}{m_s^\oplus}+\left(1-\frac{2n_q}{n}\right)\frac{m}{m_v^\oplus}\, .
\eeqy
The identity $\Upsilon=m/m_v^\oplus$ can be directly demonstrated from Eq.~(\ref{eq:def-Bq}) and the definition (\ref{eq:isoeffmass}). This identity also holds in the Fermi-liquid theory if the Landau parameters are expressible as $f_1^{qq}=f_1(n,\eta^2) (k_{\rm F}^{(q)})^2$ (the function $f_1$ being invariant under the interchange of neutrons and protons). In the TDHF theory, the parameter $\Upsilon$ is explicitly given by 
\beqy
\Upsilon = \frac{m}{m_v^\oplus}= 1 + \frac{2}{\hbar^2}\left(\frac{\delta E^j_{\rm nuc}}{\delta X_0}-\frac{\delta E^j_{\rm nuc}}{\delta X_1}\right) \rho \, .
\eeqy
This result is quite general and is applicable to any nuclear-energy density functional that depends on the nucleon densities $n_q(\pmb{r},t)$, kinetic densities $\tau_q(\pmb{r},t)$ and momentum densities $\pmb{j_q}(\pmb{r},t)$. The fact that the determinant $\Upsilon$ of the entrainment matrix is related to the isovector effective mass is not unexpected since both quantities characterize similar phenomena, namely relative motions between neutrons and protons. 

\begin{figure}[ht]
\begin{center}
\includegraphics[width=0.75\textwidth]{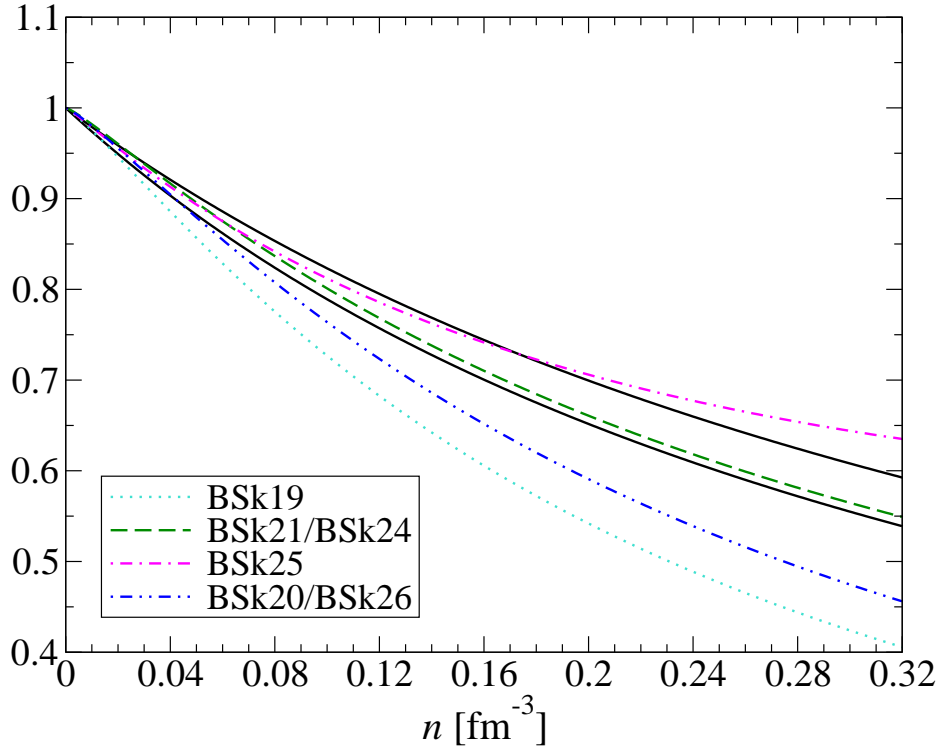}
\caption{(Color online) Variation of the isovector effective mass $m_v^\oplus/m$ with density $n$ in nucleon matter for the extended Skyrme functionals BSk19, BSk20, BSk21, BSk24, BSk25, and BSk26 ~\cite{goriely2010,goriely2013}. The upper and lower black solid lines are results from the LNS~\cite{cao2006} and Sk$\chi m^*$~\cite{zhanglim2018} functionals, which were fitted to calculations based on extended Brueckner-Hartree-Fock approach and chiral effective field theory respectively. }
\label{fig:isoeffmass_BSk}
\end{center}
\end{figure}

\begin{figure}[ht]
\begin{center}
\includegraphics[width=0.75\textwidth]{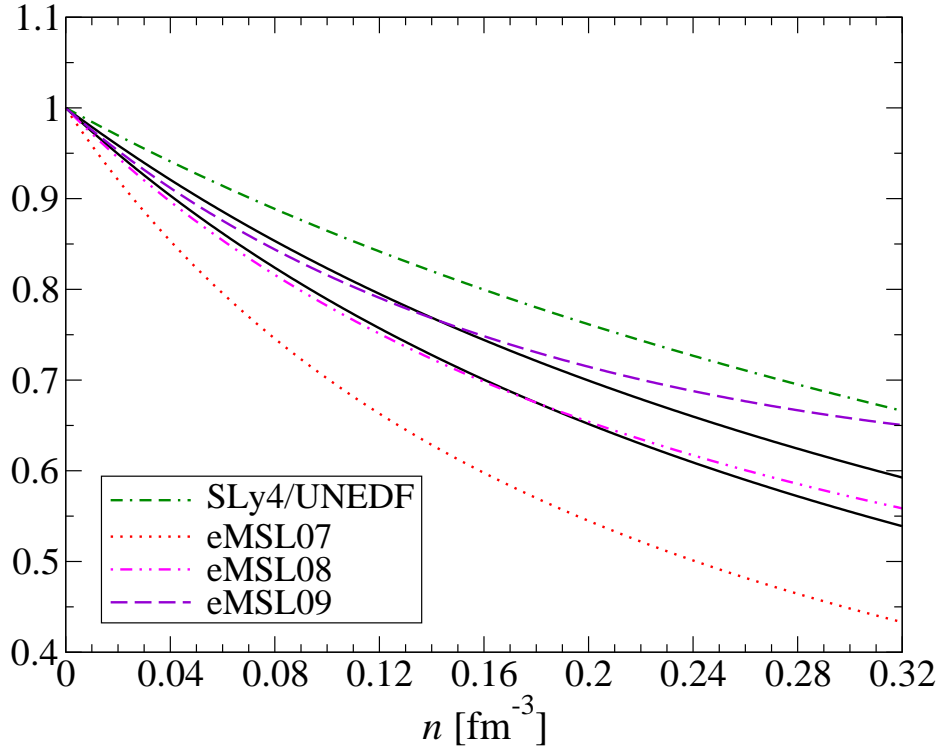}
\caption{(Color online) Same as Fig.~\ref{fig:isoeffmass_BSk} for the extended Skyrme functionals eSML07, eSML08, and eSML09~\cite{zhang2016}. For comparison, predictions from the standard Skyrme functionals SLy4~\cite{chabanat98,chabanat98err} and UNEDF~\cite{kortelainen2012} are also shown.}
\label{fig:isoeffmass_SLyEMSL}
\end{center}
\end{figure}

In principle, the isovector effective mass can be extracted from measurements of isovector giant dipole resonances in finite nuclei (the isovector effective mass being closely related to the enhancement factor $\kappa$ of the energy-weighted sum rule $m_1$). However, the values inferred from such analyses are model-dependent (see, e.g. Refs.~\cite{oishi2016,zhangchen2016}). Alternatively, the isovector effective mass can be indirectly estimated from functionals fitted to various nuclear data, as in Ref.~\cite{malik2018}. In particular, the fit to essentially all nuclear masses seems to favor values between $m_v^\oplus/m\sim0.6$ and $m_v^\oplus/m\sim0.8$ at saturation  density~\cite{goriely2010}. Considering different analyses, current estimates of isovector effective mass at saturation lie in the range $m_v^\oplus/m\sim 0.6-0.9$. These values are consistent with those found in microscopic calculations (see, e.g. Ref.~\cite{baoli2018} for a recent review). Applications to neutron stars require the knowledge of the isovector effective mass at densities ranging from about $\sim 0.08$~fm$^{-3}$ (crust-core transition) up to several times saturation density. The variations of the isovector effective mass with density as predicted by functionals LNS~\cite{cao2006} and Sk$\chi m^*$~\cite{zhanglim2018} are shown in Fig.~\ref{fig:isoeffmass_BSk}. These two functionals were directly fitted to microscopic calculations based on the extended Brueckner-Hartree-Fock approach for the former and on chiral effective field theory for the latter. These results are compared to those obtained using the Brussels-Montreal functionals~\cite{goriely2010,goriely2013}. These functionals have been employed to calculate a series of equations of state of dense matter in all regions of a neutron star in a unified and thermodynamically consistent way~\cite{potekhin2013,pearson2018}. These functionals, which were derived from extended Skyrme effective interactions containing terms that are both momentum and density dependent (see Appendix~\ref{app:Skyrme}), were precision fitted to all measured masses of nuclei with $Z,N\geq8$ from the Atomic Mass Evaluation with root-mean square deviations $\sim0.5-0.6$ MeV. These functionals were simultaneously adjusted to other experimental and theoretical nuclear data including the neutron-matter equations of state, as obtained from many-body calculations using realistic nucleon-nucleon potentials. The isovector effective masses obtained with BSk19 and BSk26 are found to be significantly smaller than the microscopic results of Refs.~\cite{cao2006,zhanglim2018}. Interestingly, these functionals are also disfavored by astrophysical observations~\cite{fantina2013,pearson2018}. On the other hand, the functionals BSk21, BSk24, and BSk25 are consistent with microscopic predictions. For comparison, results from other extended and standard Skyrme functionals, developed for astrophysical applications, are shown in Fig.~\ref{fig:isoeffmass_SLyEMSL}. The eMSL functionals~\cite{zhang2016} lead to predictions that are similar to those of the BSk series. The eMSL08 and eMSL09 parametrizations appear to yield more realistic isovector effective masses than eMSL07. The isovector effective masses obtained with the standard Skyrme functionals SLy4~\cite{chabanat98,chabanat98err} and UNEDF~\cite{kortelainen2012} are substantially higher than the microscopically calculated ones. 

The parametrisation~(\ref{eq:matrix-param1})-(\ref{eq:matrix-param3}) of the entrainment matrix is particularly well-suited for practical applications since $\Upsilon$ is independent of the composition for Skyrme-like functionals and is merely given by (see Appendix~\ref{app:Skyrme})
\beqy
\Upsilon = 1 + \frac{2}{\hbar^2}\left(C_0^\tau-C_1^\tau\right) \rho \, ,
\eeqy
where $C_0^\tau$ and $C_1^\tau$ are constant parameters for standard Skyrme functionals, and are functions of the density $n$ for the extended Skyrme functionals discussed above~\cite{krewald1977,farine2001,chamel2009,goriely2010,zhang2016}. Explicit formulas for these coefficients are given in Appendix~\ref{app:Skyrme}. 

The entrainment matrix calculated from standard Skyrme effective interactions is found to coincide with that obtained earlier using the Fermi-liquid expression~(\ref{eq:entrainment-matrix-FL}) with corresponding Landau parameters $\mathcal{F}_1^{q q^\prime}$ and effective masses $m_q^\oplus$~\cite{chamel2006}. This stems from the fact that the mass currents $\pmb{\rho_q}$ depend linearly on the superfluid velocities $\pmb{V_q}$ (the entrainment matrix is independent of $\pmb{V_q}$). However, this may not be necessarily the case for more complicated nuclear-energy density functionals. In particular, the exact expression~(\ref{eq:entrainment-matrix-TDHF}) will differ from the Fermi-liquid approximation whenever the nuclear energy functional contains terms that are not simply proportional to the fields $X_0$ and $X_1$. Examples of such functionals have been proposed in Ref.~\cite{carlsson2008}.

\section{Conclusions}

We have derived exact expressions for the local nucleon mass currents $\pmb{\rho_q}(\pmb{r},t)$ 
at any position $\pmb{r}$ and time $t$ in a cold neutron-proton mixture directly from the TDHF equations 
without any further approximation. We have also shown how to relate the spatially 
averaged mass currents to the group velocities of single-particle quantum states, demonstrating in this way the equivalence between TDHF theory and previous analyses based on the Fermi liquid approximation. Our expressions are 
very general and are applicable to both homogeneous and inhomogeneous nuclear systems. 

Focusing on the core of a neutron star, we have shown that the neutron-proton entrainment matrix can be 
conveniently expressed in terms of its dimensionless determinant $\Upsilon$, whose deviation from unity 
measures the importance of entrainment effects. This quantity depends solely on the nucleon number density $n$ and is found to coincide with the inverse of the isovector effective mass. This formulation thus allows to relate entrainment phenomena in neutron stars to isovector giant dipole resonances in finite nuclei. We have calculated the isovector effective mass for various semi-local nuclear energy-density functionals. These include the precision-fitted Brussels-Montreal functionals, for which unified equations of state of neutron stars have been already calculated~\cite{potekhin2013,pearson2018}. Comparing results to those obtained from microscopic calculations, the functionals BSk24 and BSk25 appear to be particularly well-suited for dynamical simulations of superfluid neutron stars.

\appendix
\section{Coordinate-space formulation of TDHF}
\label{app:TDHF}

Following the general definition of the density matrix, 
\begin{equation}
n_q(\pmb{r}, \sigma; \pmb{r^\prime}, \sigma^\prime;t) = \langle\Psi(t)|c_q(\pmb{r^\prime},\sigma^\prime)^\dagger c_q(\pmb{r},\sigma)|\Psi(t)\rangle\, ,
\end{equation}
where $|\Psi(t)\rangle$ is the many-nucleon wave function at time $t$, $c_q(\pmb{r},\sigma)^\dagger$ and $c_q(\pmb{r},\sigma)$ are the creation and destruction operators for nucleons of charge type $q$ at position $\pmb{r}$ with spin $\sigma$, the coordinate-space and discrete-basis representations are related by 
\beqy\label{eq:dens-matrix}
n_q(\pmb{r},\sigma;\pmb{r^\prime},\sigma^\prime;t)=\sum_{i,j} n_q^{ij}(t)\, \varphi^{(q)}_i(\pmb{r},\sigma)\varphi^{(q)}_j(\pmb{r^\prime},\sigma^\prime)^*
\eeqy
\beqy
n_q^{ij}(t)=\sum_{\sigma,\sigma^\prime}\int {\rm d}^3\pmb{r} {\rm d}^3\pmb{r^\prime}\, n_q(\pmb{r},\sigma;\pmb{r^\prime},\sigma^\prime;t) \varphi^{(q)}_i(\pmb{r},\sigma)^*\varphi^{(q)}_j(\pmb{r^\prime},\sigma^\prime)\, ,
\eeqy
denoting by $\varphi^{(q)}_i(\pmb{r},\sigma)$ the single-particle basis wavefunctions.
Making use of the completeness relations 
\beqy
\sum_i \varphi^{(q)}_i(\pmb{r},\sigma)^*\varphi^{(q)}_i(\pmb{r^\prime},\sigma^\prime)=\delta(\pmb{r}-\pmb{r^\prime})\delta_{\sigma\sigma^\prime}\, , 
\eeqy
the TDHF equations~(\ref{eq:TDHF-trad}) can thus be alternatively written as 
\beqy\label{eq:TDHF-coord-space}
\mathrm{i}\hbar \frac{\partial n_q(\pmb{r}, \sigma; \pmb{r^\prime}, \sigma^\prime;t)}{\partial t}=\sum_{\sigma^{\prime\prime}}\int\,\mathrm{d}^3\pmb{r^{\prime\prime}}\,\biggl[h_q(\pmb{r},\sigma;\pmb{r^{\prime\prime}},\sigma^{\prime\prime};t)n_q(\pmb{r^{\prime\prime}}, \sigma^{\prime\prime}; \pmb{r^\prime}, \sigma^\prime;t)\nonumber \\ -
n_q(\pmb{r},\sigma;\pmb{r^{\prime\prime}},\sigma^{\prime\prime};t)h_q(\pmb{r^{\prime\prime}}, \sigma^{\prime\prime}; \pmb{r^\prime}, \sigma^\prime;t)
\biggr]\, ,
\eeqy
with the Hamiltonian matrix defined by 
\beqy
h_q(\pmb{r},\sigma;\pmb{r^\prime},\sigma^\prime;t)=\sum_{i,j} h_q^{ij}(t)\, \varphi^{(q)}_i(\pmb{r},\sigma)\varphi^{(q)}_j(\pmb{r^\prime},\sigma^\prime)^*\, .
\eeqy
In cases for which the energy $E$ is a functional of local densities and currents, the Hamiltonian matrix can be calculated as 
\beqy
\label{eq:Hamiltonian-matrix}
h_q^{ij}(t)=\int \mathrm{d}^3\pmb{r}\,\Biggl[  \frac{\delta E}{\delta n_q(\pmb{r},t)}\frac{\partial n_q(\pmb{r},t)}{\partial n_q^{ji}(t)}+\frac{\delta E}{\delta\tau_q(\pmb{r},t)}\frac{\partial \tau_q(\pmb{r},t)}{\partial n_q^{ji}(t)}+ \frac{\delta E}{\delta\pmb{j_q}(\pmb{r},t)}\frac{\partial \pmb{j_q}(\pmb{r},t)}{\partial n_q^{ji}(t)} \Biggr]\, .
\eeqy
Using Eqs.~(\ref{eq:dens-def}), (\ref{eq:kin-def}), (\ref{eq:mom-def}) and (\ref{eq:dens-matrix}) in (\ref{eq:Hamiltonian-matrix}), and integrating by parts, the Hamiltonian matrix can be written in the form 
\beqy
h_q^{ij}(t)=\sum_{\sigma,\sigma^\prime}\int {\rm d}^3\pmb{r} {\rm d}^3\pmb{r^\prime}\,  \varphi^{(q)}_i(\pmb{r},\sigma)^*\varphi^{(q)}_j(\pmb{r^\prime},\sigma^\prime)h_q(\pmb{r},\sigma;\pmb{r^\prime},\sigma^\prime;t)\, , 
\eeqy
\beqy\label{eq:Hamiltonian-coord-space1}
h_q(\pmb{r},\sigma;\pmb{r^\prime},\sigma^\prime;t)=h_q(\pmb{r},t)\delta(\pmb{r}-\pmb{r^\prime})\delta_{\sigma\sigma^\prime}\, ,
\eeqy
with the Hamiltonian operator $h_q(\pmb{r},t)$ defined by Eq.~(\ref{eq:Hamiltonian}). From the Hermiticity property $h_q^{ij}=(h_q^{ji})^*$, we have
\beqy\label{eq:Hamiltonian-coord-space2}
h_q(\pmb{r},\sigma;\pmb{r^\prime},\sigma^\prime;t)=h_q(\pmb{r^\prime},t)^*\,\delta(\pmb{r}-\pmb{r^\prime})\delta_{\sigma\sigma^\prime}\, .
\eeqy
Note that the order of the factors in Eqs.~(\ref{eq:Hamiltonian-coord-space1}) and (\ref{eq:Hamiltonian-coord-space2}) matters: the Hamiltonian operates only on the Dirac distribution. Inserting Eqs.~(\ref{eq:Hamiltonian-coord-space1}) and (\ref{eq:Hamiltonian-coord-space2}) in (\ref{eq:TDHF-coord-space}) leads to Eq.~(\ref{eq:TDHF}).

\section{Nuclear energy-density functionals and Skyrme effective interactions}
\label{app:Skyrme}

Nuclear-energy density functionals can be obtained from the HF method using extended Skyrme effective interactions of the form
\beqy\label{eq:skyrme}
v(\pmb{r}_{i},\pmb{r}_{j}) & = & t_0(1+x_0 P_\sigma)\delta({\pmb{r}_{ij}})+\frac{1}{2} t_1(1+x_1 P_\sigma)\frac{1}{\hbar^2}\left[p_{ij}^2\,
\delta({\pmb{r}_{ij}}) +\delta({\pmb{r}_{ij}})\, p_{ij}^2 \right]\nonumber\\
&+&t_2(1+x_2 P_\sigma)\frac{1}{\hbar^2}\pmb{p}_{ij}\cdot\delta(\pmb{r}_{ij})\,\pmb{p}_{ij}+\frac{1}{6}t_3(1+x_3 P_\sigma)n(\pmb{r})^\alpha\,\delta(\pmb{r}_{ij})
\nonumber\\
&+& \frac{1}{2}\,t_4(1+x_4 P_\sigma)\frac{1}{\hbar^2} \left[p_{ij}^2\,
n({\pmb{r}})^\beta\,\delta({\pmb{r}}_{ij}) +
\delta({\pmb{r}}_{ij})\,n({\pmb{r}})^\beta\, p_{ij}^2 \right] \nonumber\\
&+&t_5(1+x_5 P_\sigma)\frac{1}{\hbar^2}{\pmb{p}}_{ij}\cdot n({\pmb{r}})^\gamma\,\delta({\pmb{r}}_{ij})\, {\pmb{p}}_{ij}  \nonumber\\
& +&\frac{\rm i}{\hbar^2}W_0(\pmb{\hat\sigma_i}+\pmb{\hat\sigma_j})\cdot
\pmb{p}_{ij}\times\delta(\pmb{r}_{ij})\,\pmb{p}_{ij} \, , 
\eeqy
where $\pmb{r}_{ij} = \pmb{r}_i - \pmb{r}_j$, $\pmb{r} = (\pmb{r}_i + 
\pmb{r}_j)/2$, $\pmb{p}_{ij} = - {\rm i}\hbar(\pmb{\nabla}_i-\pmb{\nabla}_j)/2$
is the relative momentum, $\pmb{\hat\sigma_i}$ and $\pmb{\hat\sigma_j}$ are Pauli spin matrices, 
 $P_\sigma$ is the two-body spin-exchange operator, and $n(\pmb{r})$ denotes the average nucleon number 
density. The terms proportional to $t_4$ and $t_5$ are absent in standard Skyrme functionals. Although the use of effective interactions imposes some restrictions on the form of the functional, it guarantees the cancellation of self-interaction errors~\cite{chamel2010} (nonetheless, the functional may still be contaminated by \emph{many-body} self-interactions errors, see, e.g. Ref.~\cite{duguet2014} and references therein). Parameters are usually determined by fitting various experimental and theoretical nuclear data. 

The nuclear energy is expressible as $E_{\rm nuc}=\int{\rm d}^3\pmb{r}\,\mathcal{E}_{\rm Sky}(\pmb{r})$. The 
nuclear terms contributing to the mass currents take a very simple form 
\beqy
\mathcal{E}^j_{\rm Sky}=C_0^\tau X_0 +C_1^\tau X_1\, , 
\eeqy
where the coefficients $C_0^\tau$ and $C_1^\tau$ are given by~\cite{chamel2009}
\beqy
C_0^\tau(n) = \frac{3}{16}t_1+\frac{1}{4} t_2\left( \frac{5}{4}+x_2\right)+\frac{3}{16}t_4 n^\beta+\frac{1}{4} t_5\left( \frac{5}{4}+x_5\right) n^\gamma
\eeqy
\beqy
C_1^\tau(n) =-\frac{1}{8} t_1\left(\frac{1}{2}+x_1\right)+\frac{1}{8}
t_2\left(\frac{1}{2}+x_2\right)-\frac{1}{8} t_4\left(\frac{1}{2}+x_4\right)n^\beta+\frac{1}{8}
t_5\left(\frac{1}{2}+x_5\right) n^\gamma \, .
\eeqy
The coefficients $C_0^\tau$ and $C_1^\tau$ coincide with the functional derivatives of the $E_{\rm nuc}^j$ with respect to $X_0$ and $X_1$ respectively, i.e. 
\beqy
\frac{\delta E^j_{\rm nuc}}{\delta X_0}=C_0^\tau \, , \hskip0.5cm \frac{\delta E^j_{\rm nuc}}{\delta X_1}=C_1^\tau \, .
\eeqy.

\section{Group velocity in translationally invariant systems}
\label{app:group-vel}

In nuclear systems with some translational symmetry (this includes the crystalline crust and the homogeneous core of a neutron star), the single-particle wave functions are given by Bloch waves~\cite{chamel2012}
\beqy\label{eq:Bloch-wave}
\varphi_{\pmb{k}}^{(q)}(\pmb{r},\sigma)=\frac{1}{\sqrt{V}}\exp(\mathrm{i}\, \pmb{k}\cdot\pmb{r}) \chi(\sigma) \sum_{\pmb{G}}\widetilde{\varphi}^{(q)}_{\pmb{k}}(\pmb{G}) \exp(\mathrm{i}\,\pmb{G}\cdot\pmb{r})\, ,
\eeqy
where $\pmb{G}$ are reciprocal lattice vectors and $\chi(\sigma)$ denotes the Pauli spinor. 
The HF equations (\ref{eq:HF}) can thus be written as 
\beqy
\sum_{\pmb{G^\prime}}  \widetilde{h}_{\pmb{k}}^{(q)}(\pmb{G},\pmb{G^\prime})\widetilde{\varphi}^{(q)}_{\pmb{k}}(\pmb{G^\prime})= \varepsilon_{\pmb{k}}^{(q)} \widetilde{\varphi}^{(q)}_{\pmb{k}}(\pmb{G})\, ,
\eeqy
\beqy\label{eq:Bloch-Hamiltonian}
 \widetilde{h}_{\pmb{k}}^{(q)}(\pmb{G},\pmb{G^\prime}) =\frac{1}{V}\int\mathrm{d}^3\pmb{r}\, e^{-\mathrm{i}(\pmb{k}+\pmb{G})\cdot\pmb{r}} h_q(\pmb{r}) e^{\mathrm{i}(\pmb{k}+\pmb{G^\prime})\cdot\pmb{r}}\, .
\eeqy
Making use of the normalization of the wave functions 
\beqy
\sum_{\pmb{G}}\, |\widetilde{\varphi}^{(q)}_{\pmb{k}}(\pmb{G})|^2=1\, , 
\eeqy
the single-particle energy is given by
\beqy
\varepsilon_{\pmb{k}}^{(q)}=\sum_{\pmb{G},\pmb{G^\prime}}\,  \widetilde{\varphi}^{(q)}_{\pmb{k}}(\pmb{G})^*  \widetilde{h}_{\pmb{k}}^{(q)}(\pmb{G},\pmb{G^\prime}) \widetilde{\varphi}^{(q)}_{\pmb{k}}(\pmb{G^\prime})\, .
\eeqy
According to the Hellmann-Feynman theorem~\cite{feynman1939}, we have
\beqy\label{eq:group-vel-Bloch}
\frac{1}{\hbar}\pmb{\nabla_k} \varepsilon_{\pmb{k}}^{(q)}=\frac{1}{\hbar}\sum_{\pmb{G},\pmb{G^\prime}}\,  \widetilde{\varphi}^{(q)}_{\pmb{k}}(\pmb{G})^* \biggl[\pmb{\nabla_k} \widetilde{h}_{\pmb{k}}^{(q)}(\pmb{G},\pmb{G^\prime})\biggr] \widetilde{\varphi}^{(q)}_{\pmb{k}}(\pmb{G^\prime})\, .
\eeqy
Using Eq.~(\ref{eq:Bloch-Hamiltonian}), it can be easily seen that Eq.~(\ref{eq:group-vel-Bloch}) coincides with the general definition (\ref{eq:velocity-op}), thus demonstrating 
\beqy\
\frac{1}{\hbar}\pmb{\nabla_k} \varepsilon_{\pmb{k}}^{(q)}=\pmb{v_k^{(q)}}\, .
\eeqy

In the limit of homogeneous nucleon matter as in the core of a neutron star,  $\widetilde{\varphi}^{(q)}_{\pmb{k}}(\pmb{G})=1$ for $G=0$ and $\widetilde{\varphi}^{(q)}_{\pmb{k}}(\pmb{G})=0$ otherwise, i.e. 
the single-particle wave functions reduce to plane waves
\beqy\label{eq:plane-wave}
\varphi_{\pmb{k}}^{(q)}(\pmb{r},\sigma)=\frac{1}{\sqrt{V}}\exp(\mathrm{i}\, \pmb{k}\cdot\pmb{r}) \chi(\sigma)\, .
\eeqy
In this case, the single-particle energy and the velocity can be readily calculated. Substituting Eq.~(\ref{eq:plane-wave}) in Eq.~(\ref{eq:HF}) yields 
\beqy\label{eq:s-p-energy}
\varepsilon_{\pmb{k}}^{(q)}=\frac{\hbar^2 k^2}{2m_q^\oplus} + U_q+\pmb{k}\cdot \pmb{I_q}\, .
\eeqy
Differentiating leads to 
\beqy
\pmb{v_k^{(q)}}=\frac{1}{\hbar}\pmb{\nabla_k} \varepsilon_{\pmb{k}}^{(q)} =  \frac{\hbar \pmb{k}}{m_q^\oplus}+\frac{\pmb{I_q}}{\hbar}\, . 
\eeqy

\begin{acknowledgments}
N.C. acknowledges financial support from the Fonds de la Recherche Scientifique (Belgium) under grant No. CDR-J.0115.18. This work was also partially supported by the COST action CA16214. This work was completed at the Aspen Center for Physics, which is supported by National Science Foundation grant PHY-1607611.
\end{acknowledgments}

\bibliography{references.bib}

\end{document}